# Coherent magnetic plasmon modes in a contacting gold nano-sphere chain on a gold Slab

K. N. Chen, H. Liu[*], S. M. Wang, Y. J. Zheng, C. Zhu, Y. Wang and S. N. Zhu

*Department of Physics, National Laboratory of Solid State Microstructures, Nanjing University, Nanjing 210093, People's Republic of China*

[*]*liuhui@nju.edu.cn* ;
URL: *http://dsl.nju.edu.cn/mpp*

**Abstract:** A coupled magnetic resonator waveguide, composed of a contacting gold nanosphere chain on a gold slab, is proposed and investigated. A broadband coherent magnetic plasmon mode can be excited in this one dimensional nanostructure. By employing the Lagrangian formalism and the Fourier transform method, the dispersion properties of the wave vector and group velocity of the magnetic plasmon mode are investigated. Small group velocity can be obtained from this system which can be applied as subwavelength slow wave waveguides.

## 1. Introduction

The optical properties of plasmonic structures have been widely investigated in the past two decades, concomitant with the remarkable progress in the various techniques of nanomanufacturing and chemical fabrications. Given the ability of plasmon materials to manipulate photons in the subwavelength scale, these materials are applied in many important applications, such as in optical information, nonlinear optics, and biosensors, among others. Coupling effects among plasmonic nanostructures have increasingly attracted interest in recent years. [1, 2] Various coupling processes confer on plasmon systems, which behave like chemical molecules or condensed matters, various complex optical properties. A well known coupled plasmon structure is a linear chain of metallic nanoparticles. [3-8] Electric-coupled plasmonic mode can be excited and propagated along the chain by plasmonic coupling between these particles. [9-12] The dispersion properties of such a coupled mode can be tuned by changing the electric field interactions between nanoparticles, and slow group velocity can be obtained in the structure.

With the invention of a magnetic resonator, [13] the split-ring resonator (SRR), another coupled plasmonic mode can be established along the chain of SRRs through coupling interactions between these magnetic resonators. Magnetic inductive waves are established through magnetic inductive interaction between SRRs. [14-15] A stronger exchange current interaction between SRRs could introduce magnetic plasmon (MP) modes with a broader dispersion bands. [16-17] Based on another design of magnetic resonator, the slit-hole resonator, coherent MP modes can be experimentally excited in the infrared range.[18] Through changing the incident angle, the excitation frequency can be tuned continually by both electric or magnetic resonance.[19]

In this work, a kind of coupled plasmon system is proposed based on a linear chain of contacting nanospheres on a gold slab. Due to the contacting part, electrons can flow from one sphere to another and produce strong exchange current interaction between two spheres. With the help of slab, LC resonator can be formed between two spheres and slab, which can work as magnetic unit cell. We will show that a coherent MP mode can be established in such a system. This coupled plasmon caused by exchange current is different from the widely reported plasmon mode caused by electric field interaction in nanosphere chain. A Lagrangian model is used to describe the dispersion properties of wave vector and group velocity. The simulation results prove that MP modes in the proposed structure can effectively transport slow wave signal. The dispersion properties of wave vector and group velocity of these MP modes can be tuned by altering the coupling interaction between the nanospheres and the slab.

## 2. One magnetic resonator

A single unit of the structure, with two contacting gold nanospheres placed on a gold slab, is first considered. The design of the proposed structure is illustrated in Fig. 1(a). The radius of the nanosphere is 200 nm, and the thickness of the gold layer is 50 nm. These two parts are separated by a 30 nm thick dielectric layer. The Drude model is employed to describe the dispersion of gold with $\omega_p = 1.37 \times 10^{16}$ rad/s and $\gamma = 12.24 \times 10^{13}$ rad/s for infrared light. The dielectric is assumed to be glass with $\varepsilon=2.28$ and $\mu=1$. The single unit structure is excited by a dipole source located at a distance of 120 nm in front of the first sphere. To obtain the EM resonance of the single unit, a set of finite-difference time-domain calculations using a commercial software package CST Microwave Studio is employed. The local magnetic field in the gap between the two spheres is recorded during simulations. A resonance peak is detected at 92.9 THz. The obtained distribution of current density inside the gold spheres and slab at this resonance frequency is demonstrated in Fig. 1(b). The two spheres are shown to exchange current at the contact point. The excitation also simultaneously induces current on the slab surface. The entire structure can be considered as a closed equivalent LC circuit [Fig. 1(d)]. The two spheres and the slab can be regarded as inductors in series. The middle dielectric layer works as a capacitor. However, the resonant current around the closed circuit can induce a strong magnetic field in the area surrounded by the two spheres and the slab. This phenomenon makes the structure behave like a magnetic dipole $\vec{m}$. Therefore, this mode is called as the MP mode. In the simulations, the relationship between the local magnetic field and the thickness of the dielectric layer is investigated [Fig. 1(c)]. Under the same incident intensity, when the thickness of the middle layer is increased, the magnetic resonance field will be reduced. This condition is due to the fact that EM energy cannot be well confined in the space between the nanospheres and the slab when the gap is increased. More energy is leaked out resulting in reduced resonance strength. Once the bottom gold slab is removed, the MP modes become nonexistent because a closed LC circuit cannot be formed without the slab. However, once the dielectric is removed and the nanospheres will come into contact with the slab, the MP mode by LC resonance will also disappear because of the absence of a capacitor.

**3. Two coupled magnetic resonators**

Then, the resonance of three nanospheres on a slab is also investigated [Fig. 2(a)]. Given that the former structure, i.e., two nanospheres on a slab, can be considered as a single magnetic resonator, the latter, i.e., three nanospheres, can be perceived as two coupled magnetic resonators. In the simulations, the resonance and field distribution of the latter structure are investigated. The recorded local magnetic field exhibit two resonance peaks at 65.971 and 122.73 THz [Fig. 2(b)]. For the resonance at the lower frequency, the induced currents in the two LC circuits rotate in the same direction, enabling the two magnetic dipoles to oscillate in the same phase [Fig. 2(c–d)]. This mode is called symmetry mode. In contrast, at the higher frequency the induced currents in the two LC circuits rotate in opposite directions, resulting in the anti-phase oscillation of the two magnetic dipoles [Fig. 2(e–f)]. This mode is called anti-symmetry mode.

In our system, the coupling processes between magnetic units include nearest-neighbor exchange current interaction and long range magnetic field coupling. In order to give a good description of these two interactions simultaneously, a semi analytic model is developed based on the attenuated Lagrangian formalism. If $L$ and $C$ are the effective inductance and capacitance of the structure in Fig. 1(a), respectively, then the Lagrangian of this LC resonator could be expressed as $Lag = \frac{1}{2}L\dot{q}^2 - \frac{1}{2C}q^2$, where $q$ is the oscillating charge in the structure. The structure presented in Fig. 2(a) can be seen as two connected LC circuits, and its Lagrangian should be

$$Lag = \frac{1}{2}L(\dot{q}_1^2 + \dot{q}_2^2) - \frac{1}{4C}(q_1^2 + q_2^2) + M\dot{q}_1\dot{q}_2 - \frac{1}{4C}(q_1 - q_2)^2, \qquad (1)$$

where $q_1$ and $q_2$ are the oscillating charges in two LC resonators. The first term represents the kinetic energy in the inductors. The second term is the potential electric energy in the gaps under the first and the third sphere. The interaction term $M\dot{q}_1\dot{q}_2$ is caused by magneto-inductive coupling between the two magnetic resonators. The last term corresponds to the electric potential energy stored in the gap under the second sphere, which can be seen as shared capacitor by two LC resonators [Fig.

2(d) and (f)]. Considering the Ohmic dissipation $R = \frac{1}{2}\gamma(\dot{q}_1^2 + \dot{q}_2^2)$ and substituting Eq. (1) in the Euler-Lagrange equation,

$$\frac{d}{dt}(\frac{\partial Lag}{\partial \dot{q}_m}) - \frac{\partial Lag}{\partial q_m} = -\frac{\partial R}{\partial \dot{q}_m}, \quad (m=1,2) \tag{2}$$

A pair of coupled equations can be obtained as follows:

$$\ddot{\mu}_1 + \omega_0^2\mu_1 + \Gamma\dot{\mu}_1 = \frac{1}{2}\kappa_1\omega_0^2(\mu_1 + \mu_2) - \kappa_2\ddot{\mu}_2 \tag{3a}$$

and

$$\ddot{\mu}_2 + \omega_0^2\mu_2 + \Gamma\dot{\mu}_2 = \frac{1}{2}\kappa_1\omega_0^2(\mu_1 + \mu_2) - \kappa_2\ddot{\mu}_1, \tag{3b}$$

where $\mu_m = A\dot{q}_m$ (m=1, 2) is the effective magnetic dipole, and A is the cross-sectional area surrounded by induced current in LC circuit. In Eq. (3), $\omega_0^2 = 2/(LC)$ is the eigenfrequency of the single LC circuit in Fig. 1. $\Gamma = \gamma/L$ is the damping coefficient caused by ohmic loss. Eq. (3) indicates two mechanisms, namely, the exchange of conduction current coupling and the magneto-inductive coupling, described by two coefficients, $\kappa_1$ and $\kappa_2$. In an ideal circuit, $\kappa_1 = 1/2$ and $\kappa_2 = M/L$ represent the relative strength of the mutual and self inductance of a single unit, respectively. Approximating $\Gamma/2\omega_0 \ll 1$, the two eigenfrequencies are obtained from Eq. (3), $\omega_1 = \omega_0\sqrt{(1-\kappa_1)/(1+\kappa_2)}$, and $\omega_2 = \omega_0/\sqrt{1-\kappa_2}$. The MP mode at $\omega_1$ results from the symmetric resonance of two units with $\mu_1 = \mu_2$, whereas the high frequency mode $\omega_2$ results from the asymmetric $\mu_1 = -\mu_2$.

**4. One chain of coupled magnetic resonators**

The above Lagrangian model could also be extended to the chain structure shown in Fig. 3. For an infinite chain, let $q_m$ be the oscillation charge in the *m*-th unit, and considering the coupling between magnetic resonators, the Lagrangian can be expressed as

$$Lag = \sum_m \left[\frac{1}{2}L\dot{q}_m^2 - \frac{1}{4C}(q_m - q_{m+1})^2 + M\sum_n \frac{1}{n^2}\dot{q}_m\dot{q}_{m+1}\right], \quad (m=0,\pm1\pm2,\ldots; n=1,2,3\ldots). \tag{4}$$

The third term indicates the magneto-inductive coupling between the magnetic dipoles from the nearest neighboring dipoles to the farthest ones. The Ohmic dissipation for the whole structure is

$$R = \sum_m \frac{1}{2}\gamma\dot{q}_m^2. \tag{5}$$

Substituting Eq. (4) and Eq. (5) to the Euler-Lagrange equation leads to

$$\ddot{\mu}_m + \Gamma\dot{\mu}_m + \omega_0^2\mu_m = \frac{1}{2}\kappa_1\omega_0^2(\mu_{m-1} + 2\mu_m + \mu_{m+1}) - \kappa_2\sum_n\frac{1}{n^2}(\ddot{\mu}_{m-n} + \ddot{\mu}_{m+n}), \tag{6}$$

where $\Gamma$, $\omega_0^2$, $\mu_m$, and coefficient $\kappa_1$, as well as $\kappa_2$, are as previously defined. The solutions of Eq. (6) have the form of $\mu_m = \mu_0 \exp(-m\alpha d)\exp(i\omega t - imkd)$, where d is the period of the chain, and $\alpha$ is the attenuation per unit length. With $\alpha d \ll 1$ for small damping approximation, the dispersion relationship of MP mode is obtained as

$$\omega^2 = \omega_0^2 \frac{1-\kappa_1[1+\cos(kd)]}{1+2\kappa_2 \sum_n \frac{1}{n^2}\cos(nkd)}, \tag{7}$$

where $\omega_0$ is the eigenfrequency of a single unit. Considering that a larger distance between two dipoles results in a weaker interaction, only the first eight terms of the magneto-inductive coupling are considered in the following calculations.

In the simulations, a chain of contacting nanospheres, consisting of 25 linearly arranged gold nanospheres, is used as shown in Fig. 3. Excited by a dipole source at a distance of 120 nm in front of the first sphere, the magnetic field at the last nanosphere is recorded [Fig. 4 (a)]. The results show that the transmission signal is within the frequency range 0–150 THz. Based on FDTD simulation method, the magnetic field along the nanosphere chain at different frequencies can be acquired. To calculate the dispersion of MP mode, Fourier transform is used to transform the value of the magnetic field into the wave vector region in the field in ω-k space.[7] The Fourier transform follows the formula,

$$H(\omega,k) = \int H(\omega,x) e^{ikx} dx. \tag{8}$$

The Fourier transform is processed along the chain and yields the dispersion relation. The results are shown as a grey map in Fig. 4 (b). The dispersion of the MP mode is very similar to that of surface plasmon, in which the wave vector k increases with ω from 0 to 150 THz. The theoretical dispersion result based on Eq. (7) is also deduced (See dots in Fig. 4 (b)). The Lagrangian model agree with the simulation result quite well. Actually, Lagrangian model used in this work can be well generalized to include other possible coupling interaction in many new designed coupled systems in the future, such as plasmon-mechanical or plasmon-acoustical effects.

Based on the dispersion relation of the MP mode given by Eq. (7), the group velocity can be calculated as

$$V_g = \frac{\partial \omega}{\partial k} = \frac{\omega_0^2 d}{2\omega} \cdot \frac{\kappa_1 \sin(kd) \cdot \left[1+2\kappa_2 \sum_{n=1}^{\infty} \frac{1}{n^2}\cos(nkd)\right] + 2\kappa_2 \left[1-\kappa_1(1+\cos(kd))\right] \cdot \sum_{n=1}^{\infty} \frac{1}{n}\sin(nkd)}{\left[1+2\kappa_2 \sum_{n=1}^{\infty} \frac{1}{n^2}\cos(nkd)\right]^2}. \tag{10}$$

The dispersion property of group velocity is calculated and given in Fig. 5. In this equation, only the eight nearest dipole coupling are considered. The results show that the group velocity is very small ($V_g = 0.1c$) around $k = \pi/d$ ($\omega = 140 \text{THz}$). The very small group velocity of MP mode can be obtained from the designed structures.

The slow light effect has been reported in various physical systems, including atomic gases, optical fibers, photonic crystals, and plasmon systems. In the present work, the proposed structure also demonstrates dispersive slow wave effect via MP excitation in subwavelength scale. Although the spin waves in magnetic materials have many interesting properties in the microwave range, the analog of spin waves in infrared or THz region prove to be an interesting topic and may exhibit new properties. In the present work, the slow wave is caused by the coupling effect between magnetic resonators. This wave mimics the slow spin waves in infrared or THz region, which do not occur naturally. Furthermore, given that the magnetic resonator is completely designed artificially and that the coupling interaction can be tuned at will, the dispersion of slow wave effect can be controlled completely through altering the structural parameters. Then, a slow "spin" wave can be obtained at infrared or THz region.

## 5. Conclusion

In conclusion, a one-dimensional coupled magnetic resonators waveguide structure consisting of a metal nanoparticle chain, a dielectric layer, and a metal bottom layer has been proposed and investigated. At infrared frequencies, the coherent MP modes resulting from the exchange current can be excited in the structure, which transfers EM energy in subwavelength scale transmission. A semi-analytical model is established to analyze the MP mode which can be used to calculate the dispersion relation of wave vector and group velocity. By employing Fourier transform, the dispersion relation from the simulation result is obtained. The relation agrees with the proposed theoretical model quite well. The group velocity of MP modes in the proposed structure with dispersive character indicates the presence of very small group velocity and opens many new possibilities for the design and application of slow wave waveguides at infrared or THz region.

**Acknowledgement**

This work was financially supported by the National Natural Science Foundation of China (No. 11021403, 11074119, 10874081, 60990320 and 11004102), and by the National Key Projects for Basic Researches of China (No. 2010CB630703, 2012CB921501 and 2012CB933501).

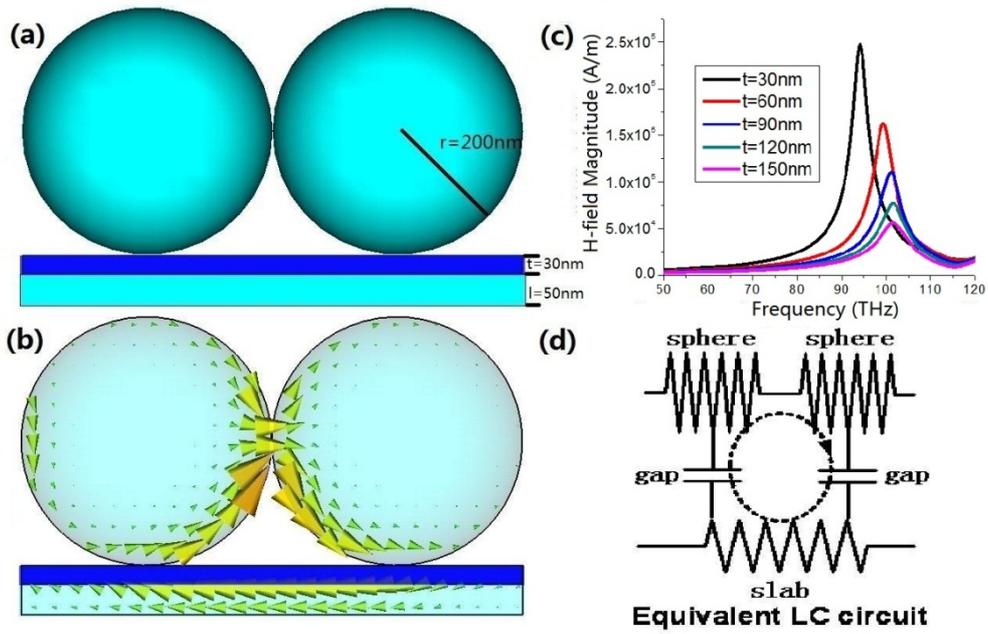

Fig.1 (a) Structure of two gold spheres on gold slab; (b) The frequency dependence of local magnetic field with different thickness of the dielectric layer; (c) The induced current at resonance frequency; (d) Equivalent LC circuits for the resonance mode.

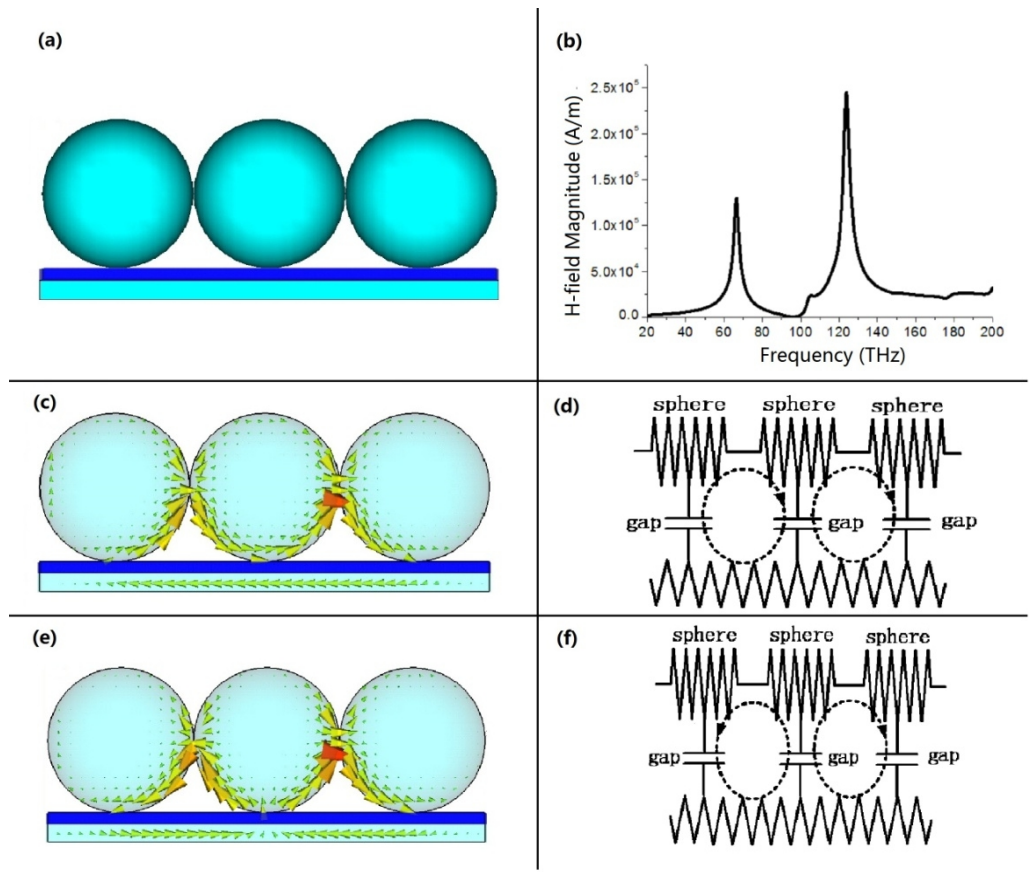

Fig.2 (a) Structure of three gold spheres on gold slab; (b) The frequency dependence of local magnetic field; The induced current distribution at lower frequency symmetry mode (c) and the equivalent LC circuit (d); The induced current distribution at higher frequency antisymmetry mode (e) and the equivalent LC circuit (f).

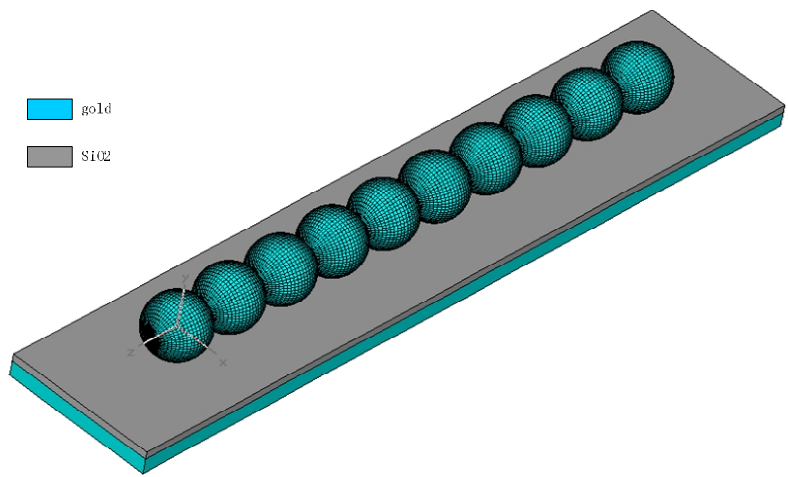

Fig.3 The structure of one dimensional chain of gold spheres on gold slab, a dipole source is used to excite the first sphere.

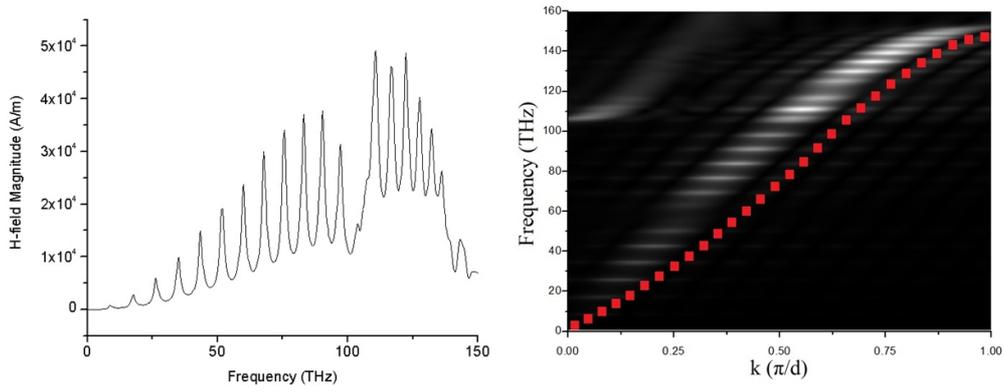

Fig.4 (a) The frequency dependence of local magnetic field at the end of chain (recorded by a probe at the last gold sphere); (b) The dispersion curve of coherent magnetic plasmon modes (grey map: simulated result; red square dot line: calculated results based on Lagrange mode).

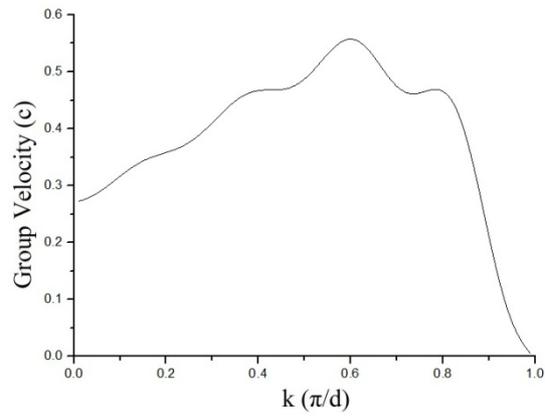

Fig.5 The calculated group velocity of the one dimensional magnetic plasmon mode.